\documentclass[prd,aps,floats,twocolumn]{revtex4}
\newcommand{\be}{\begin{equation}}
\newcommand{\ee}{\end{equation}}
\newcommand{\ba}{\begin{eqnarray}}
\newcommand{\ea}{\end{eqnarray}}
\newcommand{\bc}{}

\begin{document}

\preprint{\
\begin{tabular}{rr}
&
\end{tabular}
}
\title{Modifying gravity with the Aether: an alternative to Dark 
Matter}
\author{T.G.~Zlosnik$^{1}$, P.G.~Ferreira$^{2,3}$, and Glenn D.~Starkman
$^{4}$}
%
\affiliation{
$^1$Imperial College Theoretical Physics, Huxley Building, London SW7\\
$^{2}$Astrophysics, University of Oxford, Denys Wilkinson Building, Keble 
Road, Oxford OX1 3RH, UK\\
$^{3}$African Institute for Mathematical Sciences, 6-8 Melrose Road, Muizenberg 7945, South Africa\\
$^{4}$Department of Astronomy and Physics, Case Western Reserve, Cleveland, Ohio, U.S.A.}

\begin{abstract}
There is evidence that Newton and Einstein's theories of gravity 
cannot
explain the dynamics of a universe made up solely of baryons and radiation. To
be able to understand the properties of galaxies, clusters
of galaxies and the universe on the whole it
has become commonplace to invoke the presence of dark matter. An 
alternative
approach is to modify the gravitational field equations to accommodate
observations. We propose a new class of gravitational theories in
which we add a new degree of freedom, the Aether, in the form of
a vector field that is coupled
covariantly, but non-minimally, with the space-time metric.
We explore the Newtonian and non-Newtonian limits, discuss the 
conditions
for these theories to be consistent and explore their effect on 
cosmology.
\end{abstract}

\date{\today}
\pacs{PACS Numbers : }
\maketitle

\section{Introduction}
\noindent
 	Despite the tremendous successes of contemporary cosmology,
there is a nagging problem that refuses to go away. If we try to 
measure
the total gravitational field in the universe it far surpasses what
we would expect from the baryonic mass we can see. This is true on 
a
wide range of scales. On Kiloparsec scales
it is well known that the velocity of objects in the outer reaches of
galaxies are moving around the central core at much greater speeds 
than
what one expect from Keplerian motion due to the stars and gas.
On Megaparsec scales it has been established that the random motion of 
galaxies
in clusters is too large for these systems to remain gravitationally
bound due simply to the visible mass. And on tens to hundreds of 
Megaparsecs
there is evidence for structure in the distribution of galaxies which
should in principle have been erased by dissipational damping at 
recombination,
when the universe was a few hundred thousand years old.

 	There is a solution to this problem. One can invoke the existence
of an exotic form of matter that does not couple to light. It is cold
and clumps easily to form bound structures. The dark matter \cite{DM}
will enhance
the energy density of galaxies and clusters
and can be modeled to fit almost all observations. It will also 
sustain
gravitational potential wells through recombination and reinforce
structure on large scales. A cosmological theory based on the 
existence
of dark matter has emerged over the past twenty years with remarkable
successes and predictive power \cite{Peebles,spergel}.
Laboratory searches are under way
to find tangible evidence for dark matter candidates which go beyond
their gravitational effects.

 	One can take a different point of view. At the moment, all
evidence for dark matter comes from its dynamical effect on visible
objects. We see dark matter through its gravitational field. Could it
be that our understanding of the gravitational field is lacking? This
possibility has been mooted before. It has been proposed that the
Newton-Poisson equation, $\nabla^2\Phi=4\pi G\rho$ (where $\Phi$ is
the gravitational potential, $\rho$ is the energy density and $G$ is
Newton's constant) should be modified to
${\nabla}\cdot [f(|{\nabla \Phi}|/a_0){\nabla}\Phi]=4\pi G\rho$
where $f(x)=1$ in the strong field regime and $f(x)\simeq x$ in the 
weak field
regime. In regions of low acceleration, gravity is boosted above the 
standard
Newtonian prediction and an $f$ can be chosen to fit galactic rotation
curves \cite{MOND}. Such a theory, dubbed Modified Newtonian Dynamics
(MOND) has proven very effective and it has recently has been 
proposed that such a behaviour can emerge
from the low energy, non-relativistic limit of a fully covariant 
theory
(see \cite{Bekenstein2004,models} for
various approaches).

MOND is not without problems. It has been shown that it is less
effective at resolving the missing mass problem on the scale of
clusters of galaxies. Indeed it has been shown by Sanders \cite{Sanders} that
to correctly account for the mismatch between luminous and
dynamical mass in clusters one must invoke a small fraction of
massive neutrinos, with a mass of approximately $2$ eV. This
result has recently been reconfirmed with weak lensing data
presented by Clowe {\it et al} \cite{Clowe} and the subsequent analysis by
Angus {\it et al} \cite{Angus}.Given that
neutrinos exist, are massive and the mass required falls within the
allowed range constrained by laboratory measurements, this solution
to the missing mass problem in clusters is not outlandish. 

In this paper we show that it is possible to modify
gravity by introducing a dynamical {\it Aether} (or time-like vector 
field)
with non-canonical kinetic terms. Our proposal builds on the extensive
analysis of Einstein-Aether theories undertaken by Jacobson, 
Mattingly,
Carroll, Lim and collaborators \cite{AE},\cite{CL} and follows along
a long series of proposals by others \cite{Vector}. As the Aether vector field has a non-vanishing expectation value
it will dynamically select a preferred frame at each point in spacetime 
(i.e. the frame in which the time co-ordinate basis vector $\partial_{t}$ aligns with the direction
of the Aether field $A^{\mu}$). This violates local Lorentz invariance (and gauge invariance). Consequently, Aether theories traditionally have been used as phenomenological probes of possible Lorentz violation in quantum gravity.

As there has been recent interest in modifying gravity by using
additional scalar and vector fields it is worth initially comparing our approach to other attempts. Another group of theories retains local Lorentz invariance but introduces a vector field, the action of which breaks U(1) symmetry through a non-gauge invariant potential or kinetic term. These theories have variously been found to be able to model inflation \cite{FORD} and dark energy \cite{TRIAD}, the latter invoking a set of three identical vectors along mutually orthogonal spacial directions (the ``cosmic triad"). Such a vector field, coupled to scalar fields, has also been considered in the context of alternatives to dark matter \cite{MOFFT}. Quite distinctly, it has been 
suggested that the recent acceleration of the universe could be accounted for by allowing for non-canonical kinetic terms in the action of the electromagnetic field \cite{NONLIN}.
Recently the possibility of a scalar field coupled to the Aether as a cause of inflation has been examined \cite{SODA}. 

We will first lay out the formalism for our proposal, with the
full field equations. We will then proceed to analyze them in the
non-relativistic regime and show that it is possible to naturally 
obtain
modifications to Newtonian gravity. The physical consistency of the
theory is discussed in the weak field regime as are constraints from
the Solar system. We then briefly explore the possible impact on the
expansion of the Universe, showing that this modification
of gravity can lead to accelerated expansion at different stages
of the evolution of the universe. A specific proposal for such a 
theory
is presented and we conclude by discussing a series of open
problems.

\section{The Theory}

A general action for a vector field, $A^{\mu}$ coupled to gravity can
be written in
the form
\begin{eqnarray}
S=\int d^4x \sqrt{-g}\left[\frac{R}{16\pi G_N}+{\cal L}_{A}\right]
+S_{M} \label{genaction}
\end{eqnarray}
where $g_{\mu\nu}$ is the metric, $R$ the Ricci scalar of that metric,
$S_M$ the matter action and $\cal{L}$ is constructed to by generally
covariant and local. $S_M$ only couples to the metric, $g_{\mu\nu}$ and 
{\it not}
to $A^{\mu}$. We shall use the metric signature (-,+,+,+) throughout.

For most of this paper we will restrict ourselves consider a Lagrangian that
only depends on covariant derivatives of $A^{\mu}$ and we will consider a
that is time-like and of unit-norm. Such a theory can be written
in the form
\begin{eqnarray}
\label{eq:Lagrangian}
{\cal L}_{A}(A^{\mu},g_{\mu\nu},\lambda)&=&\frac{M^2}{16\pi G_N}
 	 {\cal F}({\cal K}) +\frac{1}{16\pi G_N}\lambda(A^\alpha A_\alpha+1)
 	\nonumber \\
{\cal K}&=&M^{-2}{\cal 
K}^{\alpha\beta}_{\phantom{\alpha\beta}\gamma\sigma}
\nabla_\alpha A^{\gamma}\nabla_\beta A^{\sigma} \nonumber \\
{\cal 
K}^{\alpha\beta}_{\phantom{\alpha\beta}\gamma\delta}&=&c_1g^{\alpha\beta}g_{\gamma\sigma}
+c_2\delta^\alpha_\gamma\delta^\beta_\sigma+
c_3\delta^\alpha_\sigma\delta^\beta_\gamma \nonumber
\end{eqnarray}
where $c_i$ are dimensionless constants and $M$
has the dimension of mass.
Note that it is possible to construct a
more complicated ${\cal K}$ by including different powers in 
$A^{\mu}$
and its derivatives. Indeed it is possible to show that Bekenstein's
theory of modified gravity \cite{Bekenstein2004} is formally 
equivalent
to a  theory with such an extended ${\cal K}$
(though with a more exotic method of achieving a
  non-vanishing vacuum-expectation value for $A^{\mu}$). We allow
for these different possibilities by deriving a general
form for the field equations below. We will comment on these models
in the discussion.

The gravitational field equations for this theory, obtained by varying $g^{\alpha\beta}$ (see \cite{CL} but also \cite{HEHL}) are
\begin{equation}
G_{\alpha\beta}=\tilde{T}_{\alpha\beta}+8\pi GT^{matter}_{\alpha\beta}
\label{fieldI}
\end{equation}
where the stress-energy tensor for the vector field is given by
\begin{eqnarray}
\tilde{T}_{\alpha\beta} &=& \frac{1}{2}\nabla_{\sigma}
({\cal F}'(J_{(\alpha}^{\phantom{\alpha}\sigma}A_{\beta)}-
J^{\sigma}_{\phantom{\sigma}(\alpha}A_{\beta)}-J_{(\alpha\beta)}A^{\sigma}))
\nonumber \\ && -{\cal F}'Y_{(\alpha\beta)}
+\frac{1}{2}g_{\alpha\beta}M^{2}{\cal F}+\lambda A_{\alpha}A_{\beta}
  \nonumber\\
{\cal F}' &=& \frac{d{\cal F}}{d{\cal K}} \nonumber \\
J^{\alpha}_{\phantom{\alpha}\sigma} &=&
(\cal{K}^{\alpha\beta}_{\phantom{\alpha\beta}\sigma\gamma}+
\cal{K}^{\beta\alpha}_{\phantom{\beta\alpha}\gamma\sigma})\nabla_{\beta}A^{\gamma}
\end{eqnarray}
Brackets around indices denote
symmetrization and $Y_{\alpha\beta}$ is the functional derivative
\begin{eqnarray}
Y_{\alpha\beta} =
\nabla_{\sigma}A^{\eta}\nabla_{\gamma}A^{\xi}
\frac{\delta(\cal{K}^{\sigma\gamma}_{\phantom{\sigma\gamma}\eta\xi})}{\delta 
g^{\alpha\beta}}
\nonumber
\end{eqnarray}
The equations of motion for the vector field, obtained by varying $A^{\beta}$ are
\begin{eqnarray}
\label{eq:veceq}
\nabla_{\alpha}({\cal F}'J^{\alpha}_{\phantom{\alpha}\beta})
+{\cal F}'y_{\beta}&=&2\lambda A_{\beta}
\label{vectoreom}
\end{eqnarray}
where once again we define the functional derivative
\begin{eqnarray}
y_{\beta}=\nabla_{\sigma}A^{\eta}\nabla_{\gamma}A^{\xi}
\frac{\delta(\cal{K}^{\sigma\gamma}_{\phantom{\sigma\gamma}\eta\xi})}
{\delta A^\beta} \nonumber
\end{eqnarray}
Variations of $\lambda$ will fix

\begin{eqnarray}
A^\mu A_\mu=-1
\end{eqnarray}
By inspection, contracting both sides of (\ref{eq:veceq}) 
with $A^{\beta}$ leads to a solution for $\lambda$ in terms of the the vector field
and its covariant derivatives. These equations allow us to study a general theory of the form presented
in equation \ref{genaction} with a time-like vector field. 
 For our particular, restricted choice
of ${\cal K}$ we have $Y_{\alpha\beta}=-c_{1}\left[ (\nabla_{\nu}A_{\alpha})(\nabla^{\nu}A_{\beta})-(\nabla_{\alpha}A_{\nu})(\nabla_{\beta}A^{\nu})\right]$ and $y_\beta=0$.

\section{The non-relativistic regime: Newtonian and MONDian limit}

Having established our general theory, we can now explore its 
properties
in various different regimes.  We start in the static, weak field,
non-relativistic limit.  We must expand both the
metric and vector field around a fixed, Minkowski space background.
We choose to use the Newtonian (or Poisson) gauge for which the metric takes the form:

\begin{eqnarray}
g_{\mu\nu}dx^{\mu}dx^{\nu} &=&  -(1+2\Phi)dt^{2}+(1-2\Psi)\delta_{ij}dx^{i}dx^{j}
\end{eqnarray}
where the functions $\Phi$ and $\Psi$ are assumed to be of first order in smallness and, to this order in perturbation theory, to depend only on the spatial coordinates $x^{i}$. Now turning to the form of the vector field, we see that in the background we may simply choose $A^{\mu}= \delta^{\mu}_{\phantom{\mu}0}$. We may then adopt the following ansatz for the perturbed $A^{\mu}$:

\begin{eqnarray}
A^{\mu}\partial_{\mu} =  (1+B^{0})\partial_{t} + B^{i}\partial_{i}
\end{eqnarray}
Due to the fixed-norm constraint upon $A^{\mu}$ we immediately have that:

\begin{eqnarray}
B^{0} =  -\Phi
\end{eqnarray}
Furthermore we assume that $B^{i}$ is of higher than first-order in smallness.
Given our assumptions regarding the perturbed form of $g_{\mu\nu}$ and $A^{\mu}$ it may consequently by shown that  to first order the $ij$th component of the Aether's stress energy tensor 
disappears. The $ij$th component of the Einstein equations then yields $\Phi=\Psi$. The equality of scalar potentials is true of General Relativity
in the absence of anisotropic stress and simplifies the equations dramatically.
The time-time component of Einstein's equations and the time component of the vector field equation then become:

\begin{eqnarray}
2\nabla^{2}\Phi  +(c_{1}-c_{3})\nabla.({\cal F}'\nabla \Phi)-
\lambda  &=& 8\pi G\rho \label{first} \\
c_{3}\nabla.({\cal F}'\nabla\Phi)&=&-\lambda \label{eq:npe}
\end{eqnarray}
The constraint fixes ${\cal K}$ to be
\begin{eqnarray}
{\cal K}= -c_{1}\frac{(\nabla\Phi)^{2}}{M^{2}}
\end{eqnarray}
Taking $c_{1}<0$ ensures that $\cal{K}$ is positive.
Using equation (\ref{eq:npe}) to solve for the Lagrange multiplier field $\lambda$ we
obtain the following equation for $\Phi$:
\begin{equation}
\label{eq:ss}
\nabla.((2+c_{1}{\cal F}')\nabla\Phi)=8\pi G\rho
\end{equation}
This is a modified Poisson equation of precisely the form proposed by Bekenstein and Milgrom.
If such a theory is to have
$\nabla.(\left|\nabla\Phi\right|\nabla\Phi)\propto \rho$
  in the limit of small $\nabla\Phi$ we require that:

\begin{equation}
\lim_{|\nabla\Phi| << M}[2+c_{1}{\cal F}'] \propto {\cal K}^{\frac{1}{2}} 
\end{equation}
Integrating then we have that
${\cal F} =  \alpha{\cal K}+\beta {\cal K}^{\frac{3}{2}}$
where $\alpha$ and $\beta$ are constants.  Hence we can
construct a theory with `MONDian' limit on galactic scales, this holding
as long as we identify $M$ with something of the order of $a_{0}$ 
\cite{MOND}.

The limit as ${\cal K}\rightarrow 0$ is worth considering in more detail. 
Considering a test particle a distance $r$ away from an isolated source, 
the Modified Poisson Equation dictates that for $|\nabla\Phi|<<M$ the 
gravitational force will vary as $1/r$. This combined with the geodesic equation
suggests that in this regime the metric component $g_{00}$ grows approximately
as $+\ln (r)$ and therefore becomes ill defined in the limit of $r\rightarrow \infty$. This has been a generic feature of past metric theories of
MOND. It was noted in \cite{WoodardNonlocal} that the value of this growing term 
will typically vary by around only $10^{-6}$ from the radius of onset of MOND 
in a system to the present Hubble radius ($> 10^{27}$cm). Therefore, the weak
field approximation is unlikely to be seen to break down. The situation becomes 
more complicated if the theory is such that MOND only arises in a cosmological 
background (for instance if $g_{\mu\nu}A^{\mu}A^{\nu}$ is not fixed and the `scale' $M$ is contingent on 
its value at the particular cosmological era). It would only be consistent then
to formulate the weak field limit in a FRW background rather than Minkowski spacetime and
the asymptotic form of the metric may be expected to differ.

The Solar System can supply us with stringent constraints on
the weak field limit of these  theories.
Accelerations are typically substantially larger than
  $M$ (again assumed to be $\sim a_{0}$) and so requiring concordance 
with observations could constrain
the possible form of ${\cal F}$ for large ${\cal K}$.
Poisson's equation $\nabla^{2}\Phi=4\pi G \rho$ is an excellent 
approximation
in the Solar System and thus we shall expect the contribution of
$c_{1}{\cal F}'$ in (\ref{eq:ss}) to be small.
We may expect then that ${\cal F}'$ in this limit can be expanded as a
power series in inverse powers of ${\cal K}^{\frac{1}{2}}$. That is:
\begin{equation}
\label{eq:limiting}
\lim_{|\nabla\Phi| >> M}[{\cal F}'({\cal K})]=
\sum_{i=1}^{\infty}\frac{\eta_{i}}{{\cal K}^{\frac{i}{2}}}
\end{equation}
where $\eta_{i}$ are constants.

Consider the leading term $\eta_{1}$/${\cal K}^{\frac{1}{2}}$.
  We would expect such a term in spherical symmetry to result in a
constant anomalous acceleration equal to
$\frac{\sqrt{-c_{1}\eta_{1}^{2}}M}{2}$.
The existence of such a term is particularly constrained
by observed bounds on the variation of Kepler's constant
$GM_{\odot}$. For instance, observations between Earth and Mars
restrict the acceleration to be less than approximately 
$10^{-9}$ms$^{-1}$
\cite{sanders}.
  We shall see that $\eta$ and $c$ values are typically of order 
unity
so the above is rather restrictive. Even under the assumption of 
spherical
symmetry, the field equations for the theory are enormously 
complicated.
The inherent nonlinearity in the weak field limit provided by
${\cal F}'$ presents a considerable challenge \cite{BMSS}.
We note however that the fixed norm constraint on $A^{\mu}$ will,
in the weak field limit, force terms of the form
$\nabla_{\mu}A^{\alpha}\nabla_{\nu}A^{\beta}$ and
$\nabla_{\mu}\nabla_{\nu}A^{\alpha}$ to be at most of the order of terms
comprising the components of the Einstein tensor $G_{\mu\nu}$.
For the limiting form of ${\cal F}'$ given in (\ref{eq:limiting})
we expect terms in the Aether stress energy tensor to be
schematically of the form
$\frac{\alpha_{i}}{{\cal K}^{\frac{i}{2}}}c_{j}
[\nabla_{\mu}\nabla_{\nu}A^{\alpha}$ and $\nabla_{\mu}A^{\alpha}\nabla_{\nu}A^{\beta}]$
and higher order derivative terms suppressed relatively by factors of
$1$/${\cal K}^{\frac{i}{2}}$.

At Mercury, the ratio $|\nabla\Phi|/ M$ is of order $10^{8}$. 
Provisionally neglecting terms in ${\cal F}'$ of ${\cal K}^{-\frac{1}{2}}$ (see 
above),
we then expect corrections to terms in $G_{\mu\nu}$ in the inner solar
system to be of order $10^{-16}c_{j}\alpha_{i} /c_{k}$.
  It is tempting to conclude that two `Parameterized Post Newtonian' (PPN) parameters
measurable by inner solar system effects, $\beta$ and $\gamma$,
would generally be expected to deviate from the predictions of
General Relativity by a similar order. The complete set of PPN
 coefficients have been obtained for the case ${\cal F}({\cal K})\propto {\cal K}$ (\cite{AE}, 3rd reference)
. In particular the coefficients describing `preferred frame' effects were found to be only consistent with experimental bounds for
specific combinations of the $c_{i}$; they are expected to be a particularly strong test of more general forms of ${\cal F}$.

 Additionally we note that the asymptotic
behaviour of ${\cal F}$ is consistent with our assumption that the term $M^{2}\frac{{\cal F}}{2}g_{\mu\nu}$ 
is second order in perturbations. We see that for ${\cal K}\ll 1$, ${M^{2}\cal F} \rightarrow |\nabla\Phi|^{2}$,
and for ${\cal K} \gg 1$, ${M^{2}\cal F}$ shall be at most $\alpha_{1}M|\nabla\Phi|$.

Finally we briefly consider the effect of allowing for first-order spatial components $A^{i}$ of the vector field.
It has been found \cite{DODLIG} that in Bekenstein's theory of modified gravity the growth of large scale structure is necessarily accompanied by growth in $A^{i}$ so it is not unreasonable to expect that such behaviour
shall be present in the model considered here. It may be readily checked that in the \emph{static} weak-field limit that an $A^{i}$ of order $\epsilon$ will only contribute to $\tilde{T}_{00}$ and 
$\tilde{T}_{ii}$ at order two in $\epsilon$ and above. However, as the numerator in ${\cal K}$ is a second order quantity, it shall generally be affected. If $A^{i}$ has only a radial component (i.e. $A^{r}\neq 0$), it may be checked that ${\cal K}$ is modified to

\begin{eqnarray}
{\cal K} = \frac{-c_{1}(\nabla\Phi)^{2}+c_{2}(\nabla_{r}A^{r})^{2}}{M^{2}} 
\end{eqnarray}
where the covariant derivative $\nabla_{r}A^{r}$ is of order two or greater in smallness. We shall
see later that $c_{2}$ is preferably the same sign as $c_{1}$. The effect of gradients of the radial component of the vector field then is to decrease ${\cal K}$ for a given $|\nabla\Phi |$. Recall that the onset of MONDian behaviour coincides with ${\cal K}<<1$. Therefore an $A^{r}(r)$ in this model will generally hasten the onset of this limit, the effect being to further increase the gravitational field for a given $\rho$ (see (\ref{eq:ss})). It requires further work to see whether this can appreciably counter
MOND's problems on the scale of clusters of galaxies.

\section{Further Constraints}

Recall that ${\cal F}$ tends to $\alpha {\cal K}+\beta {\cal K}^{\frac{3}{2}}$ for small ${\cal K}$ (i.e. far from a source). Therefore, in general, when considering classical 
perturbations
one is effectively considering a theory with ${\cal F}\sim \cal{K}$,
as in the Einstein-Aether theory with minimal couplings.
This can be used to study the consistency
of these theories in the perturbative regime.
Lim \cite{LIM} has considered the dominant term
in the limit where metric and vector field perturbations decouple.
The vector field propagates in flat spacetime and allows for a
decomposition of perturbations into spin-0 and spin-1 components.
The requirement that the perturbations can be consistently quantized,
that the spin-0 component propagates subluminally and nontachyonically
when quantized, and that metric perturbations do not propagate
superluminally place the following restrictions on the $c_{i}$:

\begin{eqnarray}
c_{1} &<& 0 \\
c_{2} &\leq& 0\\ 
c_{1}+c_{2}+c_{3} &\leq& 0.
\end{eqnarray}
Additional constraints can be obtained via astroparticle physics. 
The observation of ultra-high energy cosmic rays implies a lack of energy 
loss via gravitational Cherenkov radiation. 
This radiation is expected when gravitational waves propagate subluminally. 
With the Aether, the usual transverse traceless modes exist along with three
coupled Aether-metric modes. The speeds of propagation of each mode
are functions of the $c_{i}$ and have been calculated in \cite{AE} for perturbations in Minkowski spacetime. The squared-speed $s^{2}_{tt}$ of the 
transverse-traceless mode is: 

\begin{eqnarray}
s^{2}_{tt}=1/(1+(c_{1}+c_{3})) 
\end{eqnarray}

and the squared-speeds of the transverse Aether and trace Aether-metric modes ($s^{2}_{tA}$ and $s^{2}_{tr}$ respectively)
are given by:

\begin{eqnarray}
s^{2}_{tA} &=& \frac{(c_{1}+\frac{c_{1}^{2}}{2}-\frac{c_{3}^{2}}{2})}{(c_{1}(1+c_{13}))}\\
s^{2}_{tr} &=& \frac{(c_{123}/c_{1})(2+c_{1})}{[2(1-c_{2})^{2}+c_{123}(1-c_{2}-c_{123})]} 
\end{eqnarray}

where $c_{ijk..}=c_{i}+c_{j}+c_{k}+..$ and $c_{i}$ are the opposite sign to those considered in \cite{AE}.

Computing the expected degree of gravitational Cherenkov radiation for the above modes and comparing this with observational bounds, it has been found \cite{ELL} that for the case ${\cal F}({\cal K})={\cal K}$ that extremely stringent 
constraints could be placed on the $c_{i}$. For instance, they may be satisfied when the $c_{i}$ are mutually related such that the metric and Aether-metric modes propagate at precisely the speed of light. If this is not the case then the magnitudes 
of the $c_{i}$ must be severely diminished. It is expected that such an analysis could powerfully constrain
more general combinations of ${\cal F}$,$c_{i}$, though any nonlinearity is expected to complicate 
the results when propagation is considered against a curved background spacetime.

Note that such an analysis is by no means complete. It has been
shown that a very restricted class of Einstein-Aether theories do
not have positive Hamiltonians and therefore are inherently unstable
at both the classical and quantum levels \cite{clayton}. Furthermore we 
are
considering non-linear functions of ${\cal K}$ and hence instabilities
may arise in non trivial backgrounds. A more detailed analysis of
individual cases for ${\cal F}$ is needed yet first indications
are that these theories are healthy.

\section{Cosmology}

This class of modified theories are generally covariant. This gives us
the possibility of exploring their properties on large scales and
in particular we can consider the case of a  homogeneous and
isotropic universe in which the metric is of the form
$ds^2= -{dt}^{2}+a(t)^{2}\delta_{ij}dx^{i}dx^{j}$ where $t$ is cosmic proper time and 
$a(t)$ is
the scale factor.
The vector field must respect the spatial homogeneity and isotropy of 
the system and so will only have a non-vanishing `$t$' component; the 
constraint
fixes $A^{\mu} = (1,0,0,0)$. The energy-momentum tensor of the
matter is of the form ${T}^{matter}_{\alpha\beta} = \rho 
U_{\alpha}U_{\beta}
  +P(g_{\alpha\beta} + U_\alpha U_\beta)$ where
$\rho$ is the energy density, $P$ is pressure
and we have introduced a four-vector $U^{\mu}$ which satisfies
$g_{\mu\nu}U^{\mu}U^{\nu} = -1$. We may additionally then fix $U^{\mu}=(1,0,0,0)$. Given these forms, we find that:
\begin{eqnarray}
\nabla_{\mu}A^{\mu}&=& 3H\nonumber \\
{\cal K} &=& 3\frac{\alpha H^{2}}{M^{2}} \nonumber
\end{eqnarray}
where $H \equiv \frac{\dot{a}}{a}$,
the dot denotes differentiation with respect to $t$, and,
following \cite{CL}, we define $\alpha = c_{1}+3c_{2}+c_{3}$.

Note that now ${\cal K}$ is negative, unlike the non-relativistic limit encountered 
above.
This means that the dynamics of static, spherically symmetric systems
on the one hand
and of relativistic cosmologies on the other probe
completely different branches of ${\cal F}$. The modified Einstein's equations now become:
\begin{eqnarray}
[1-{\cal F}'\alpha]H^{2}+\frac{1}{6}{\cal F}M^{2} = \frac{8\pi 
G}{3}\rho
\label{00} \nonumber \\
-[1-2{\cal F}'\alpha]H^{2}-2[1-\frac{1}{2}{\cal F}'\alpha]\frac{\ddot{a}}{a}+\dot{{\cal F}'}\alpha H-\frac{1}{2}{\cal F}M^{2}= 8\pi G P\label{trace} \nonumber
\end{eqnarray}
These can in fact be rewritten in a more useful form:
\begin{eqnarray}
\left[1-\alpha{\cal K}^{1/2}\frac{d}{d{\cal K}}\left(\frac{\cal F}{{\cal K}^{1/2}}\right)\right]H^2 &=& \frac{8\pi G}{3}\rho \label{00m} \\
\frac{d}{dt}(-2H+{\cal F}'\alpha H)&=&8 \pi G (\rho+P) \label{summ}
\end{eqnarray}
Once again these equations are general but we can see in their 
structure,
  interesting possibilities. If we take ${\cal F}=0$, we recover the 
standard
cosmology. We can do this by either setting it to $0$ or choosing 
${\cal K}$
to have $c_1=-c_3$ and $c_2=0$ so that it becomes the Maxwell tensor. 
We
  then have a theory which will modify gravity on galactic
and super-galactic scales but leaves the expansion of the universe 
unchanged.

It is interesting to explore the possibility that the vector field
may affect the late time expansion of the Universe. From Einstein's equations,
we can see that for it to behave like a cosmological constant, 
$\Lambda$ we
must have
\begin{equation}
\label{eq:c}
-{\cal K}{\cal F}'+\frac{1}{2}{\cal F}=\frac{\Lambda}{M^2} 
\end{equation}
which we can solve to find:
${\cal F}=\beta (- {\cal K})^{1/2}-\frac{2\Lambda}{M^2}$. 

Where $\beta$ is a constant of integration. Now 
note that for
positive arguments of ${\cal F}$
  we have been equating the mass scale $M$ with the
acceleration scale $a_0\simeq 10^{-10}$ms$^{-2}\simeq cH_0$ where
$c$ is the speed of light and $H_0$ is the Hubble constant. If we 
assume
that is also true for negative values, we find that ${\cal F}$ is of 
order unity. I.e. there is a natural
relation between the scale of the cosmological constant and the 
fundamental
mass scale in our theory. A choice of such asymptotic behaviour
close to ${\cal K}=0$ does introduce an undesirable discontinuity in
${\cal F}$ at the origin; however, as the behaviour in (\ref{eq:c})
will lead to a constant ${\cal K}$ (as $\dot{H}=0$),this 
form of ${\cal F}$ need
only hold in the domain $|{\cal K}/3\beta|>\sim H_0$ and does not
inform the desirable choice of ${\cal F}$ around ${\cal K}=0$.
Accelerated expansion can be reached in a more natural way
by positing
\begin{eqnarray}
{\cal F}=\gamma (-{\cal K})^n \nonumber
\nonumber
\end{eqnarray}
 for a certain range of ${\cal K}$
For arbitrary $n$ Equation (\ref{00}) becomes
\begin{eqnarray}
\left[1+\epsilon\left(\frac{H}{M}\right)^{2(n-1)}\right]H^2
=\frac{8\pi G}{3}\rho 
\nonumber
\end{eqnarray}
where $\epsilon=-(1-2n)\gamma(-3\alpha)^{n}/6$.For an
appropriate choice of $\gamma$ and $n$ we have $\epsilon<0$
and we find that $H$ tends to an attractor 
\begin{eqnarray}
H\rightarrow H_{eq}=M(-\epsilon)^{1/2(1-n)}
\nonumber
\end{eqnarray}
This regime is approached asymptotically with
\begin{eqnarray}
H-H_{eq}\simeq H_{eq}^{-1}\left(\frac{8\pi G\rho}{3}\right) \nonumber
\end{eqnarray}
i.e. as $a^{-3}$ in the matter era. This gives us a particularly 
elegant, dynamical mechanism for approaching accelerated
expansion without invoking a cosmological constant. A small
scale is still invoked but it is naturally related to
the acceleration scale needed to trigger the onset of the MONDian
regime on galactic scales.

\section{Discussion}

The discussion throughout this paper has been rather general. We have 
neither
chosen a specific form of ${\cal K}$ or ${\cal F}$ although
we have constrained the asymptotic form of the latter. It is
instructive to pick a simple example.
If we choose $c_{1}=-1$, $c_{2}=0$, and $c_{3}=1$ we will
recover the canonical form ${\cal K}\propto F_{\alpha\beta}F^{\alpha\beta}$
where $F_{\mu\nu}\equiv 2\partial_{[\mu}A_{\nu ]}$. 
A possible functional form for ${\cal F}$ is then:
\begin{eqnarray}
{\cal F}({\cal K})=4({\cal K}^{\frac{1}{2}}-\ln({\cal 
K}^{\frac{1}{2}}+1)) \nonumber
\end{eqnarray}
From which  we recover the modified Newton-Poisson equation:
\begin{equation}
\nabla. \left( \frac{\frac{|\nabla\Phi|}{M}}{\frac{|\nabla\Phi|}{M}+1} 
\nabla \Phi\right)=4\pi G \rho
\end{equation}
With an appropriate choice of the value of $M$ we recover the field
equation considered in \cite{BinneyFamaey}, which was found to give a
satisfactory fit to the terminal velocity curve of the Milky Way.
As noted, the choice $c_{1}=-c_{3},c_{2}=0$ as here forces ${\cal K}$ 
to
vanish in the case of spatial homogeneity and isotropy.
Therefore with this choice of ${\cal F}$ the Aether has no influence 
on
cosmological background evolution. The inclusion of a nonzero $c_{2}$ 
modifies the kinetic term of the metric and may have significant effects
both in the static, weak-field limit and the cosmological background.

There are a number of theoretical and phenomenological issues
that remain to be addressed.

For a start, the presence of the non-dynamical Lagrange-multiplier 
field $\lambda$
in the Lagrangian (\ref{eq:Lagrangian}) is perhaps unappealing.
Its sole role is to impose the constraint that $A^\mu$ is 
unit-timelike.
The same can be accomplished by replacing the $\lambda$-term in the
Lagrangian by a potential term $-V(A^\mu)$, by dint of which $A^\mu$
at low energy acquires a vacuum expectation value such that $A^\mu 
A_\mu=-1$.
For example  $V(A^\mu) = \frac{1}{2}\mu^4 \left(A^\mu A_\mu +1 
\right)^2$,
with $\mu$ a constant with dimensions of mass.
At energy scales below $\mu$, one expects that indeed $g_{\mu\nu}A^{\mu}A^{\nu}=-1$,
and that ``radial'' excitation of $A^{\mu}$
(i.e. of $a \equiv \mu A$)
will have positive mass-squared  $m_a^2\simeq \mu^2$.
So long as $\mu^2$ is sufficiently large, the low-energy
phenomenology of this model should be identical to that
of (\ref{eq:Lagrangian}).
A more exotic possibility is to construct more complex ${\cal K}$
which have minima for time-like $A^{\mu}$. Indeed the theory
proposed in \cite{Bekenstein2004} is of this form \cite{ZFS1}.

Closely related to the previous point is the fundamental origin
of such an Aether. We have kept the discussion general in the
hope that a more fundamental theory of fields or strings may
pin down the form of ${\cal K}$ and ${\cal F}$. Indeed,
the possibility of Lorentz violating vector fields has cropped
up in attempts to extend the standard model of particle physics.
Most notably it has been argued that such a field may arise
in higher dimensional theories as a low energy by product of
string theory \cite{KOSTEL}. It would be interesting
to explore the range of current candidates for
low energy string theory to find a possible candidate for such an 
Aether field.
What is clear is that a Lagrangian of the form we require will {\it 
not} appear
from the standard perturbative approach to constructing effective
field theories as in \cite{gripaios}. Non-perturbative effects
must come into play.

We have endeavoured to explore some of the observational constraints.
Preliminary indications are that these theories are compatible with
Solar System constraints. A more detailed analysis is needed with
complete spherically symmetric solutions that can then be used to
calculate the Post Newtonian Parameters \cite{will}. We have also
shown that, as yet there is sufficient freedom in the choice of
${\cal F}$ to obtain different cosmological behaviours, from no
effect to early or late time acceleration. The next step is to
follow in the footsteps of \cite{SMFB} and calculate the evolution
of linear perturbations. A priori it is unclear whether perturbations
in the Aether will have the same effect as dark matter in sustaining
perturbations through the Silk damping regime during recombination
\cite{PY}. Indeed this may be
the most stringent test such theories have to pass to be
viable alternatives to dark matter.

\newpage 
{\it Acknowledgments}:
We thank P. Candelas, C.Skordis, and A. Mozaffari for helpful discussions.
TGZ is supported by a PPARC studentship. GDS is supported by 
Guggenheim
and Beecroft Fellowships.


\vspace{-.3in}
\begin{thebibliography}{99}
\vspace{-.7in}

\bibitem{DM} J.F.~Navarro {\it et al}, Astroph. J, 462, 563 (1996);
M.~Davis {\it et al}, Astroph. J, 292, 371, (1985)
;G.~Bertone, D.~Hooper and J.~Silk, Phys. Rep. 405, 279 
(2005).
\bibitem{Peebles} P.J.~Peebles, Astroph. J., 248, 885 (1981).
\bibitem{spergel} D.N.~Spergel {\it et al}, {\tt astro-ph/0603449}
\bibitem{MOND} M.Milgrom, Astroph. J., 270, 365 (1983); ibid., 
Astroph.
J., 270, 371 (1983); ibid., Astroph. J., 270, 384 (1983).
\bibitem{Bekenstein2004}J.D.~Bekenstein,  Phys.Rev. \textbf{D70} 083509 (2004)
\bibitem{models}R.H.~Sanders,Mon.Not.Roy.Astron.Soc. 363,459 
(2005); I.~Navarro and K.~Van~Acoleyen, gr-qc/0512109; P.~Mannheim, Int.J.Mod.Phys.
\textbf{D12} 893-904(2003)
\bibitem{Sanders}R.H. Sanders, astro-ph/0212293 
\bibitem{Clowe}D. Clowe {\it et al}, astro-ph/0608407 
\bibitem{Angus} G.W. Angus {\it et al}, astro-ph/0609125
\bibitem{AE}T.~Jacobson and D.~Mattingly, Phys. Rev. \textbf{D64}, 024028
(2001)
;  T.~Jacobson and D.~Mattingly, Phys. Rev. \textbf{D70}, 024003 (2004)
; B.Z.~Foster and T.~Jacobson, Phys. Rev. \textbf{D73} 064015 (2006)
;  C.~Eling Phys. Rev. \textbf{D73} 084026 (2006)
;  B.Z.~Foster Phys.Rev. \textbf{D73} 104012 (2006)
\bibitem{Vector} C.M.~Will and K.~Nordvedt, Astroph. J., 177, 775 
(1972);
R.W.~Hellings and K.~Nordvedt, Jr.,
  Phys. Rev. \textbf{D7}, 3593 (1973); V.A. Kostelecky and S. Samuel,
Phys. Rev. \textbf{D40}, 1886 (1989); B.Basset {\it et al}, Phys. Rev. \textbf{D62}, 
103518
(2000)
\bibitem{FORD} L.H.~Ford, Phys. Rev. \textbf{D40}, 967 (1989)
\bibitem{TRIAD} C.~Armendariz-Picon, JCAP \textbf{0407}, 007 (2004)
\bibitem{MOFFT} J.~Moffat, JCAP \textbf{0603}, 004 (2006) 
\bibitem{NONLIN} M.~Novello {\it et al} Phys. Rev. \textbf{D69}, 127301 (2004) 
\bibitem{SODA} S.~Kanno and J.~Soda, Phys. Rev. \textbf{D74}, 063505 (2006) 
\bibitem{CL} S.M.~Carroll and E.A.~Lim, Phys. Rev. \textbf{D70}, 123525 (2004)
\bibitem{HEHL} C.~Heinicke,P.~Baekler, and F.W.~Hehl Phys.Rev. \textbf{D72}, 025012 (2005)
\bibitem{gripaios} B.M.~Gripaois, JHEP \textbf{10}, 069 (2004).
\bibitem{sanders} R.H.~Sanders, astro-ph/0602161
\bibitem{BMSS} J.~Bekenstein and J.Magueijo Phys. Rev. \textbf{D73}, 103513 (2006) 
\bibitem{BinneyFamaey} J.J.~Binney and B.~Famaey 
Mon.Not.Roy.Astron.Soc. 363 (2005)
\bibitem{DODLIG}S.~Dodelson and M.~Liguori,  Phys.Rev.Lett. \textbf{97}, 231301 (2006) 
\bibitem{LIM} E.A.~Lim, Phys. Rev. \textbf{D71}, 063504 (2005)
\bibitem{ELL}
  J.~W.~Elliott, G.~D.~Moore and H.~Stoica, JHEP {\bf 0508},066 (2005) 
\bibitem{WoodardNonlocal} M.E.~Soussa and R.P.~Woodard, Class.Quant.Grav. 20 (2003)
\bibitem{clayton} M.Clayton {\tt gr-qc/0104103}
\bibitem{ZFS1} T.~Zlosnik, P.G.~Ferreira and G.~Starkman, {\tt 
gr-qc/0606039}
\bibitem{tevesperturb}C.Skordis, ArXiV astro-ph/0511591
\bibitem{KOSTEL} V.A.Kostelecky, Phys. Rev. \textbf{D69}, 105009 (2004)
\bibitem{will} C.M.~Will, {\it gr-qc/0510072}
\bibitem{SMFB} C.Skordis, D.Mota, P.G.~Ferreira and C.~Boehm, Phys. 
Rev. Lett.,
\textbf{96}, 011301 (2006).
\bibitem{PY}P.J.~Peebles and J.T.~Yu, Astroph. J. 162, 170 (1970); 
M.L.Wilson
and J.Silk, Astroph. J. 243, 12 (1981)
\end{thebibliography}
\end{document}